\documentclass[a4paper, 12pt]{article}
\usepackage{latexsym}
\usepackage{amssymb}
\usepackage{amsfonts} 
\usepackage{amsmath}
\usepackage[mathscr]{euscript}

\author{
Michael Loss\footnotemark[1] ,  Tadahiro Miyao\footnotemark[2] and Herbert  Spohn\footnotemark[3]\\ 
\footnotemark[1] {\it Georgia Institute of Technology, School of Mathematics,}\\
{\it Atlanta, Georgia 30332-0160}\\
\texttt{loss@math.gatech.edu}\\
\footnotemark[2]
\footnotemark[3]  {\it Zentrum Mathematik,}
{\it Technische Universit\"at M\"unchen,}\\ 
{\it  D-85747 Garching, Germany}\\
\footnotemark[2]  \texttt{miyao@ma.tum.de},\ 
\footnotemark[3]  \texttt{spohn@ma.tum.de}}

\title{\textsf{Kramers degeneracy theorem  in nonrelativistic QED}}
\date{\empty}

\newcommand{\one}{{\mathchoice {\rm 1\mskip-4mu l} {\rm 1\mskip-4mu l}
{\rm 1\mskip-4.5mu l} {\rm 1\mskip-5mu l}}}
\newcommand{\h}{\mathfrak{h}}
\newcommand{\Hil}{\mathcal{H}}
\newcommand{\ex}{\mathrm{e}}
\newcommand{\D}{\mathrm{dom}}

\newcommand{\im}{\mathrm{i}}
\newcommand{\Fock}{\mathfrak{F}}

\newcommand{\la}{\langle}
\newcommand{\ra}{\rangle}

\newcommand{\BbbR}{\mathbb{R}}

\newcommand{\BbbC}{\mathbb{C}}
\newcommand{\vepsilon}{\varepsilon}
\newcommand{\vphi}{\varphi}

\newcommand{\A}{A_{\kappa,K}}

\newcommand{\kl}{k, \lambda}

\def\Sumoplus{\sideset{}{^{\oplus}_{n\ge 0}}\sum}

\begin{document}

\newtheorem{define}{Definition}[section]
\newtheorem{Thm}[define]{Theorem}
\newtheorem{Prop}[define]{Proposition}
\newtheorem{lemm}[define]{Lemma}
\newtheorem{rem}[define]{Remark}
\newtheorem{assum}{Condition}
\newtheorem{example}{Example}
\newtheorem{coro}[define]{Corollary}

\maketitle

\begin{abstract}
Degeneracy of the  eigenvalues of the Pauli-Fierz Hamiltonian with spin $1/2$ is proven by the Kramers degeneracy theorem.
The Pauli-Fierz Hamiltonian at fixed total momentum is also investigated.
\end{abstract}

\section{Introduction}
Among the fundamental observations in physics the anomalous Zeeman effect must be high on anybody's list. Due to the spin of
the electron the degeneracy of energy levels is lifted by the interaction with an external magnetic field. On the other hand in zero external field, there is no splitting of the energy levels observed, despite the fact that electrons
carry their magnetic radiation field with them. In purely physical terms the total system, i.e., the radiation included, has a time reversal symmetry which can be used to explain that energy levels must be degenerate. This is important
for understanding the spectrum of atoms but also justifies the use of effective spin Hamiltonians.
It is the aim of this little note to explain all this on the basis of non-relativistic QED using the Kramers degeneracy theorem. 

We consider an electron coupled to the quantized radiation field described what is sometimes called the Pauli-Fierz Hamiltonian.
The existence of a ground state for non-relativistic QED is by now rather well understood. 
The reader may consult \cite{ BFP,BFS, LLG,  Hiroshima1, LiebLoss} for the use of various techniques. In the absence of spin the ground state
can be shown to be  unique, either by estimating the overlap between ground states and the Fock vacuum \cite{BFS} or relying on 
the Perron-Frobenius theorem via the functional integral formula for the heat kernel \cite{Hiroshima1}.

If spin is included, the ground state is no longer unique.  Exact double degeneracy of the ground state was first proven by Hiroshima and Spohn \cite{HS}. They prove first that the eigenvalue is at least doubly degenerate by a perturbative argument and then show that
the degeneracy cannot be more than two by calculating the overlap with the Fock vacuum as in \cite{BFS}. Both calculations require that the fine
structure constant (for a given ultra violet cutoff) is sufficiently small.

 In this note we give a simple proof of the at least double degeneracy based on the Kramers degeneracy theorem which first appeared in \cite{MS}.
In view of its importance of the result and the simplicity of the argument we decided to present this argument in a separate
paper. The reader will see that our proof clarifies an  essential structure behind Hiroshima and Spohn's proof. 
It should be remarked that not only ground states but also all eigenvectors, should they exist, are at least doubly degenerate. 
As far as we know, no one has applied the Kramers theorem for this purpose  to nonrelativistic QED before.
Our arguments also apply to $N$-electron system coupled to the Maxwell field provided $N$  is odd.  

The Hamiltonian $H$ of an electron interacting with the radiation field is translation invariant and hence the total momentum is conserved.
Thus $H$ can be written as a direct integral $H=\int^{\oplus}_{\BbbR^3} H(P)\, \mathrm{d}P$, where $H(P)$ is the Hamiltonian with  a fixed total momentum $P$.  The existence of the ground state of $H(P)$
is established by \cite{BCF, JFroehlich1, Spohn2} under suitable conditions. Exact double degeneracy of it was shown by Hiroshima and Spohn 
 \cite{HS}, too. Their proof is a modification of the proof of the corresponding statement for $H$. Kramers degeneracy theorem applies to this case 
 as well and yields the degeneracy of every eigenvalue.

 Recently Hiroshima gave another proof of the ground state
 degeneracy by using the fact that $H(P)$ commutes with rotations \cite{Hiroshima3}. (We remark that, his method is motivated by Sasaki \cite{sasaki}.)
  This argument, however, depends on the choice of the polarization vectors of the quantized radiation field. This makes the arguments somewhat 
  complicated. Our method is  free from the choice of  the polarization vector. Hiroshima's proof, however,  clarifies the symmetry property 
  of the ground states.
 
 This paper is organized as follows. In section 2 we establish an abstract framework of the Kramers degeneracy theorem.
 We apply this abstract theory to the Pauli-Fierz Hamiltonian with spin $1/2$ in section 3 and to the Pauli-Fierz Hamiltonian at fixed total momentum
 in section 4.  In section 5 we remark about  the $N$-electron system coupled to the radiation field.
 Section 6 is devoted to discussion of Hiroshima-Spohn's lemma.

\begin{flushleft}{\bf Acknowledgements}\\
 This work was  supported  by the DFG under the grant SP181/24. M.L. would like to acknowledge partial support by NSF grant DMS-0653374.
 \end{flushleft} 
 
\section{Abstract theory}

\subsection{Reality preserving operators}
Let $\h$ be a complex Hilbert space  and $j$ be an involution on
$\h$. Namely (i) $j$ is antilinear, (ii) $j^2=\one$, the identity on
$\h$, (iii) $\|jx\|=\|x\|$ for all $x\in \h$.
Let $\h^j=\{x\in\h\, |\, jx=x\}$. Then $\h^j$ is a real Hilbert space. 
A vector $x$ in $\h$ is said to be {\it $j$-real } if $x\in\h^j$ holds.  

 A linear operator $a$ on
$\h$ is called to be {\it reality preserving} with respect to $j$ if $j\D(a)\subseteq \D(a)$
(equivalently $j\D(a)=\D(a)$) and 
\[
 ajx=jax
\] 
for all $x\in\D(a)$. ¡¡We remark that   $a$ preserves reality w.r.t. $j$ if and only if
$a\h^j\cap \D(a)\subseteq \h^j$ holds.
We denote the set of all reality preserving operators w.r.t. $j$ by
$\mathfrak{A}_j(\h)$.  Basic  property of $\mathfrak{A}_j(\h)$ is stated as below.
\begin{Prop}
$\mathfrak{A}_j(\h)$ is a real algebra, namely, we have the following.
\begin{itemize}
\item[{\rm (i)}] $\forall a,b\in\mathfrak{A}_j(\h)$ $\forall \alpha,\beta\in \BbbR$, $\alpha a+\beta b\in \mathfrak{A}_j(\h)$. 
\item[{\rm (ii)}] $\forall a,b\in\mathfrak{A}_j(\h)$, $ab\in \mathfrak{A}_j(\h)$ provided the product is defined.
\end{itemize}
\end{Prop}
{\it Proof.} This is an easy exercise. $\Box$

\subsection{Abstract Kramers degeneracy theorem}
The following proposition is an abstract version of the Kramer's degeneracy, found in the physical literature.
\begin{Prop}\label{Kramer}
Let $\vartheta$ be an antiunitary operator with $\vartheta^2=-\one$. Let $H$ be a self-adjoint operator. Assume that $H$
commutes with $\vartheta$. Then each eigenvalue of $H$ is at least doubly degenerate.
\end{Prop}
{\it Proof.} 
Let $\vphi$ be an eigenvector for the eigenvalue $E$. Since $H$ commutes with $\vartheta$,
 one sees
 \[
 H\vartheta \vphi=\vartheta H\vphi=E\vartheta \vphi.
 \]
 Hence $\vartheta \vphi$ is also eigenvector for $E$. By the antiunitarity, $\la \vartheta \psi_1, \vartheta \psi_2\ra=\la \psi_2, \psi_1\ra$
 for all $\psi_1, \psi_2$. Thus, using $\vartheta^2=-\one$, 
 \begin{align*}
 -\la \vphi, \vartheta \vphi\ra=\la \vartheta (\vartheta \vphi), \vartheta \vphi\ra=\la \vphi, \vartheta \vphi\ra.
 \end{align*}
 Hence $\la \vphi, \vartheta \vphi\ra=0$.
 $\Box$
 \medskip\\
 
 The following lemma is a direct consequence of the functional calculus.
 
 \begin{Prop}\label{Functional}
 Assume that $H$ and  $\vartheta$ satisfy the conditions in Proposition \ref{Kramer}. Let $f$ be a real-valued measurable  function on $\BbbR$.
 Then $f(H)$ defined by the functional calculus commutes with $\vartheta$ too.  
 \end{Prop}

\subsection{Kramers degeneracy in the system  with spin $1/2$}
Let us consider a direct sum Hilbert space
$\Hil=\h\oplus\h$. Let $j$ be an involution  on $\h$. Then $J=j\oplus j=\left(\begin{matrix}
j&0\cr
0&j\cr
\end{matrix}
\right)\ $
is an involution  on $\Hil$ too. Let $\sigma_1, \sigma_2, \sigma_3$ be the
$2\times2$ Pauli matrices on $\Hil$:
\[
\sigma_1=\left(\begin{matrix}
0&\one\cr
\one&0\cr
\end{matrix}
\right),\ 
\sigma_2=\left(
\begin{matrix}
0&-\im\cr
\im&0\cr
\end{matrix}
\right),\ 
\sigma_3=\left(
\begin{matrix}
\one&0\cr
0&-\one\cr
\end{matrix}
\right).
\]

We restrict our attention
to the following case. Let $H_0$ be a semibounded self-adjoint operator on $\Hil$
having a form
\[
 H_0=\left(
\begin{matrix}
A&0\cr
0&A\cr
\end{matrix}
\right).
\]
Hence $A$ is self-adjoint and  bounded from below.
We assume the following.
\begin{flushleft}
{\bf (H.1)}\ \  $A\in \mathfrak{A}_j(\h)$.
\end{flushleft}
Clearly each eigenvalue of $H_0$ is doubly degenerate. We consider the
perturbation by a symmetric operator $H_{\mathrm{I}}$ of the form:
\begin{align*}
H_{\mathrm{I}}=\sigma\cdot B=\sum_{j=1}^3\sigma_i B_j=\left(
\begin{matrix}
B_3&B_1-\im B_2\cr
B_1+\im B_2&-B_3\cr
\end{matrix}
\right),
\end{align*}
where each $B_i,\ i=1,2,3$ is a symmetric operator on $\h$ possessing
the following properties:
\begin{flushleft}
{\bf (H.2)}\ \  Each $B_i$ is infinitesimally small with respect to  $A$.\\
{\bf (H.3)}\ \  $\im B_i\in \mathfrak{A}_j(\h),\, i=1,2,3$.
\end{flushleft}
Remark that (H.3) is equivalent to $jB_ix=-B_i jx$ for all $x\in\D(B_i),
\\ i=1,2,3$.
The condition (H.2) guarantees the self-adjointness of the following operator
\[
 H=H_0+H_{\mathrm{I}}.
\]
Define an antiunitary operator by
\begin{align}
 \vartheta=\sigma_2 J.\label{CommuteJ}
\end{align}
Then $\vartheta$ is an antiunitary operator satisfying $\vartheta^2=-\one$. 

\begin{Thm}\label{Abstract} 
Under the assumptions (H.1), (H.2) and (H.3) each eigenvalue of $H$ is at least
 doubly degenerate. 
\end{Thm} 
{\it Proof.} Noting the  facts $\vartheta
\sigma_i=-\sigma_i\vartheta,\  j B_i=-B_i j,\ i=1,2,3$ and the assumption (H.1), one has 
\begin{align*}
\vartheta H_0&=H_0\vartheta,\\
\vartheta \sigma\cdot B&=\sigma\cdot B\vartheta
\end{align*}
which implies $\vartheta H=H\vartheta$. Hence, by Proposition  {\ref{Kramer}}, each eigenvalue of $H$ is at least
 doubly degenerate. $\Box$



\section{Pauli-Fierz Hamiltonian with spin $1/2$}
The Pauli-Fierz Hamiltonian is given by 
\begin{align*}
H_{\mathrm{PF}}=\frac{1}{2}\big(-\im\nabla_x+eA(x)\big)^2+\frac{e}{2}\sigma\cdot B(x)+V(x)+H_{\mathrm{f}}
\end{align*}
acting in $L^2(\BbbR^3;\BbbC^2)\otimes\Fock$, where $\Fock$ is the photon Fock
space
\[
 \Fock=\Sumoplus L^2(\BbbR^3\times\{1,2\})^{\otimes_{\mathrm{s}}n},
\]
  $\h^{\otimes_{\mathrm{s}}n}$ means the $n$-fold symmetric tensor product of $\h$ with the convention $\h^{\otimes_{\mathrm{s}}0}=\BbbC$.
The quantized vector potential $A(x)=(A_1(x), A_2(x), \\A_3(x))$ is given
by 
\[
 A(x)=\sum_{\lambda=1,2}\int_{|k|\le
 \Lambda}\frac{\mathrm{d}k}{\sqrt{2(2\pi)^3|k|}}\vepsilon(\kl)\Big(\ex^{\im
 k\cdot x}a(\kl)+\ex^{-\im k\cdot x}a(\kl)^*\Big),
\]
where $\vepsilon(\kl)$ is a polarization vector which is real valued and
measurable, $\Lambda$ is the ultraviolet cutoff. Here
$a(\kl), a(\kl)^*$ are the annihilation
and creation operators which satisfy the standard commutation relations
\begin{align*}
 [a(\kl), a(q,\mu)^*]&=\delta_{\lambda\mu}\delta(k-q),\\ [a(\kl),a(q,\mu)]&=0=[a(\kl)^*,a(q,\mu)^*].
\end{align*}
$B(x)$ is the quantized magnetic field defined by
\begin{align*}
&B(x)\\
&=\mathrm{rot} A(x)\\
&=\im \sum_{\lambda=1,2}\int_{|k|\le
 \Lambda}\frac{\mathrm{d}k}{\sqrt{2(2\pi)^3|k|}}(k\times \vepsilon(\kl))\Big(\ex^{\im
 k\cdot x}a(\kl)-\ex^{-\im k\cdot x}a(\kl)^*\Big).
\end{align*}
$H_{\mathrm{f}}$ is the field energy given by 
\[
 H_{\mathrm{f}}=\sum_{\lambda=1,2}\int_{\BbbR^3}\mathrm{d}k\, |k|a(\kl)^*a(\kl).
\]
Throughout this section, we assume the following:
\begin{flushleft}
{\bf (V)} \ \ $V$ is infinitesimally small with respect to $-\Delta_x$.
\end{flushleft}
Then, by \cite{HH, Hiroshima2}, $H_{\mathrm{PF}}$ is self-adjoint on
$\D(-\Delta_x)\cap \D(H_{\mathrm{f}})$, bounded from below.

Our Hilbert space $L^2(\BbbR^3;\BbbC^2)\otimes\Fock$ is naturally
identified with  
\[
 L^2(\BbbR^3; \Fock)\oplus L^2(\BbbR^3; \Fock).\]
 Under this identification, $H_{\mathrm{PF}}$ is
understood as follows:
\begin{align*}
H_{\mathrm{PF}}&=H_0+\frac{e}{2}\sigma\cdot B(x),\\
H_0&=\left(\begin{matrix}
H_{\mathrm{Spinless}}&0\cr
0&H_{\mathrm{Spinless}}\cr
\end{matrix}
\right),\\
H_{\mathrm{Spinless}}&=\frac{1}{2}\big(-\im\nabla_x+eA(x)\big)^2+V(x)+H_{\mathrm{f}}.
\end{align*}
Note the following facts:
\begin{itemize}
\item[(i)] $\sigma\cdot B(x)$ is infinitesimally small w.r.t. $H_0$.
\item[(ii)] $H_{\mathrm{Spinless}}$ is self-adjoint on $\D(-\Delta_x)\cap
	   \D(H_{\mathrm{f}})$ and bounded from below by \cite{HH, Hiroshima2}.
\end{itemize}
On $L^2(\BbbR^3; \Fock)$, we take the following involution:
\begin{align*}
 j\vphi=\Sumoplus\overline{\vphi^{(n)}}(-x; k_1, \lambda_1,\dots,k_n, \lambda_n), \\  x\in \BbbR^3,\ \ (k_i, \lambda_i)\in \BbbR^3\times \{1,2\}
\end{align*}
for $\vphi=\sum_{n\ge 0}^{\oplus}\vphi^{(n)}(x; k_1, \lambda_1, \dots,
k_n,  \lambda_n)\in L^2(\BbbR^3;\Fock)$.
Since the annihilation operator $a(\kl)$ acts by
\begin{align*}
a(\kl)\vphi=\Sumoplus\sqrt{n+1}\vphi^{(n+1)}(x;k, \lambda, k_1, \lambda_1, \dots, k_n, \lambda_n)
\end{align*}
for $\vphi\in L^2(\BbbR^3; \Fock)$, one has 
\[
 ja(\kl)=a(\kl)j, \ \ ja(\kl)^*=a(\kl)^*j.
\]
Namely the annihilation and creation operators are  reality
preserving w.r.t. $j$. As a consequence, we obtain
\begin{align*}
j(-\im\nabla_x)&=(-\im\nabla_x)j,\\
jA(x)&=A(x)j,\\
jB(x)&=-B(x)j,\\
jH_{\mathrm{f}}&=H_{\mathrm{f}}j,\\
jV(x)&=V(-x)j.
\end{align*}
By the above relations, one arrives at the following:
\begin{lemm}Assume that $V(-x)=V(x)$.
\begin{itemize}
\item[{\rm (i)}] The spinless Hamiltonian $H_{\mathrm{Spinless}}$
		 preserves the reality w.r.t. $j$, equivalently
		 $H_{\mathrm{Spinless}}\in\mathfrak{A}_j(L^2(\BbbR^3
		 ; \Fock))$. This is corresponding to
		 (H.1).
\item[{\rm (ii)}]  $jB(x)=-B(x)j$, that is,  $\im B_i(x)\in
		 \mathfrak{A}_j(L^2(\BbbR^3; \Fock))$. This  corresponds to  (H.3).
\end{itemize}
\end{lemm} 
Thus we can apply Theorem \ref{Abstract} to obtain the following.

\begin{Thm}
Let $\vartheta=\sigma_2 J$ with $J=j\oplus j$.  Then $\vartheta$ is an antiunitary operator satisfying  $\vartheta^2=-\one$.
Assume that (V) holds. Moreover suppose that  $V(x)=V(-x)$. Then we obtain
\[
 \vartheta H_{\mathrm{PF}}=H_{\mathrm{PF}}\vartheta.
\]
In particular, each eigenvalue of $H_{\mathrm{PF}}$ is degenerate.
\end{Thm}

\section{Pauli-Fierz Hamiltonian at fixed total momentum}\label{Fiber}
Let us consider the   Hamiltonian at  fixed total momenutm
\begin{align*}
H_{\mathrm{PF}}(P)=\frac{1}{2}\big(P-P_{\mathrm{f}}+eA(0)\big)^2+\frac{e}{2}\sigma\cdot B(0)+H_{\mathrm{f}},\ \ P\in \BbbR^3,
\end{align*}
where $P_{\mathrm{f}}$ is the field momentum defined by
\[
 P_{\mathrm{f}}=\sum_{\lambda=1,2}\int_{\BbbR^3}\mathrm{d}k\, ka(\kl)^*a(\kl).
\]
Our Hilbert space is $\BbbC^2\otimes \Fock$. 

Under the natural identification  $\BbbC^2\otimes \Fock=\Fock\oplus
\Fock$, our Hamiltonian is represented as 
\begin{align*}
H_{\mathrm{PF}}(P)&=H_0(P)+\sigma\cdot B(0),\\
H_0(P)&=\left(\begin{matrix}
H_{\mathrm{Spinless}}(P)&0\cr
0&H_{\mathrm{Spinless}}(P)\cr
\end{matrix}
\right),\\
H_{\mathrm{Spinless}}(P)&=\frac{1}{2}\big(P-P_{\mathrm{f}}+eA(0)\big)^2+H_{\mathrm{f}}.
\end{align*}
By \cite{Hiroshima3,  LMS}, $H_{\mathrm{Spinless}}(P)$ is
positive and  self-adjoint on $\D(P_{\mathrm{f}}^2)\cap
\D(H_{\mathrm{f}})$. Moreover 
$\sigma\cdot B(0)$ is infinitesimally small with respect to $H_0(P)$. 
We choose an involution $j$ by
\[
 j\vphi=\Sumoplus \overline{\vphi^{(n)}}(k_1, \lambda_1,\dots, k_n, \lambda_n), \ \ ( k_i, \lambda_i)\in\BbbR^3\times\{1,2\}
\]
for each
$\vphi=\sum_{n\ge 0}^{\oplus} \vphi^{(n)}(k_1, \lambda_1,\dots,k_n, \lambda_n)\in
\Fock$. 
Then  the annihilation and creation operators $a(\kl),
a(\kl)^*$ are  reality
preserving w.r.t. $j$ again, because the action of $a(\kl)$ on
$\vphi=\sum_{n\ge 0}^{\oplus} \vphi^{(n)}(k_1, \lambda_1, \dots, k_n, \lambda_n) \\  \in\Fock$ is given by
\[
 a(\kl)\vphi=\Sumoplus \sqrt{n+1}\vphi^{(n+1)}(k, \lambda, 
 k_1, \lambda_1, \dots, k_n, \lambda_n).
\]
Accordingly  one can easily see that 
\begin{align*}
jA(0)&=A(0)j,\\
jB(0)&=-B(0)j,\\
jH_{\mathrm{f}}&=H_{\mathrm{f}}j,\\
jP_{\mathrm{f}}&=P_{\mathrm{f}}j,
\end{align*}
which imply that $H_{\mathrm{Spinless}}(P)$ and
$\im B_i(0), \  i=1,2,3$ are  in $\mathfrak{A}_j(\Fock)$.
Thus we can apply Theorem \ref{Abstract} and
obtain the following:

\begin{Thm}\label{Commute}
Let $\vartheta=\sigma_2 J$ with $J=j\oplus j$.  Then $\vartheta$ is an antiunitary operator with $\vartheta^2=-\one$.
Moreover  we obtain
\[
 \vartheta H_{\mathrm{PF}}(P)=H_{\mathrm{PF}}(P)\vartheta.
\]
In particular, each eigenvalue of $H_{\mathrm{PF}}(P)$ is at least doubly degenerate.
\end{Thm}

\section{$N$-electron  system with one fixed nucleus }
In this section, we  remark on  an $N$-electron system governed by the following Hamiltonian
\begin{align*}
H_N
=&\sum_{j=1}^N\Big\{ \frac{1}{2} \Big(\sigma^{(j)}\cdot \big(-\im \nabla_{x_j}+eA(x_j)\big)\Big)^2    -\frac{Ze^2}{|x_j|}      \Big\}\\
&+\sum_{1\le i<j\le N}
\frac{e^2}{|x_i-x_j|}
+H_{\mathrm{f}}.
\end{align*}
$H_N$ is acting in $\mathcal{H}_N=(L^2(\BbbR^3; \BbbC^2)^{\otimes_{\mathrm{as}}N})\otimes \Fock$ and self-adjoint on\\ $\cap_{j=1}^N\D(\Delta_{x_j})\cap\D(H_{\mathrm{f}})$
by \cite{Hiroshima2, HH}. $L^2(\BbbR^3; \BbbC^2)^{\otimes_{\mathrm{as}N}}$ means the $N$-fold antisymmetric tensor product of $L^2(\BbbR^3; \BbbC^2)$.
$\sigma^{(l)}=(\sigma_1^{(l)}, \sigma_2^{(l)}, \sigma_3^{(l)}),\ \ l=1,\dots, N$ is given by
\[
\sigma_i^{(l)}=\one\otimes\cdots\otimes \one \otimes \underset{l\, \mathrm{th}}{\sigma_i}\otimes \one\otimes \cdots \otimes \one.
\]
Each $\vphi\in \mathcal{H}_N$ is expressed as 
\[
\vphi=
\Sumoplus \vphi^{(n)}(x_1, \dots, x_N; \tau_1, \dots, \tau_N; k_1, \lambda_1, \dots, k_n, \lambda_n),
\]
where $x_i\in \BbbR^3$, $\tau_i=1,2$ and $(k_i, \lambda_i)\in \BbbR^3\times \{1,2\}$.
Let us define
\[
J\vphi=\Sumoplus \overline{\vphi^{(n)}}(-x_1, \dots, -x_N; \tau_1, \dots, \tau_N; k_1, \lambda_1, \dots, k_n, \lambda_n)
\]
and 
\begin{equation*}
\vartheta=\Pi_{l=1}^N \sigma_2^{(l)} J.     \label{TimeRe}
\end{equation*}
Clearly $\vartheta$ is antiunitary. If  $N$ is {\it odd}, then $\vartheta$ satisfies $\vartheta^2=-\one$.
Passing through  similar arguments as  in Section \ref{Fiber}, one arrives at the following.

\begin{Thm}
Let $\vartheta$ be defined as above. Then $\vartheta$ is antiunitary and one has 
\[
\vartheta H_N=H_N\vartheta.
\]
Moreover if $N$ is odd, $\vartheta^2=-\one$ holds. Hence each eigenvalue of $H_N$ is at least doubly degenerate.
\end{Thm}

\begin{rem}
{\rm We can treat the $N$-electron Hamiltonian with a fixed total momentum discussed in \cite{LMS} by the similar way.
}
\end{rem}

\section{Discussion of Hiroshima-Spohn's lemma}

Let us consider the Hamiltonian $H(P)$ in this section.
In \cite{HS} the following lemma is a key ingredient of their proof. Here we derive the lemma from our view point, because this  lemma is itself  interesting. 

\begin{lemm}{\rm  (Hiroshima-Spohn \cite{HS})}
Let $x\in \BbbC^2$. Then there exists $a(t)\in \BbbR$ independent of $x$ such that for all $t\ge 0$
\begin{align*}
\la x\otimes \Omega, \ex^{-tH(P)} x\otimes \Omega\ra=a(t)\|x\|^2_{\BbbC^2},
\end{align*}
where $\Omega=1\oplus0\oplus0\oplus \cdots\in \Fock$, the Fock vacuum.
\end{lemm}
{\it Proof.}
Since $H(P)$ commutes with $\vartheta$ defined in Theorem \ref{Commute},  $\ex^{-t H(P)}$ also commutes with $\vartheta$ for all $t\ge 0$ by Proposition \ref{Functional}.
Hence for any $x, y\in \BbbC^2$, one sees
\begin{align*}
\la x\otimes \Omega, \ex^{-tH(P)} y\otimes \Omega\ra=\la \vartheta y\otimes \Omega, \ex^{-t H(P)}\vartheta x\otimes \Omega\ra.
\end{align*}
If we choose $x=\binom{1}{0}$ and $y=\binom{0}{1}$, one obtains 
\begin{align*}
\big\la \tbinom{1}{0}\otimes \Omega, \ex^{-tH(P)}\tbinom{0}{1}\otimes \Omega \big\ra=-\big\la \tbinom{1}{0}\otimes \Omega, \ex^{-tH(P)}\tbinom{0}{1}\otimes \Omega \big\ra\end{align*}
which implies
\begin{align*}
\big\la \tbinom{1}{0}\otimes \Omega, \ex^{-tH(P)}\tbinom{0}{1}\otimes \Omega \big\ra=0.
\end{align*}
Similarly if we choose $x=\binom{1}{0}$ and $y=\binom{1}{0}$, one gets
\begin{align*}
\big\la \tbinom{1}{0}\otimes \Omega, \ex^{-tH(P)}\tbinom{1}{0}\otimes \Omega \big\ra=\big\la \tbinom{0}{1}\otimes \Omega, \ex^{-tH(P)}\tbinom{0}{1}\otimes \Omega \big\ra.
\end{align*}
Hence we have the desired result. $\Box$
\medskip\\

Similarly we can also show a slightly generalized version.
\begin{Prop}
Let $x\in \BbbC^2$ and $f\in L^{\infty}(\BbbR)$.  If $f$  is real valued, then we have 
\begin{align*}
\la x\otimes \vphi,  f(H(P)) x\otimes \vphi\ra&=\|x\|^2_{\BbbC^2}\big\la \tbinom{1}{0}\otimes \vphi, f(H(P))\tbinom{1}{0}\otimes \vphi \big\ra\\
&=\|x\|^2_{\BbbC^2}\big\la \tbinom{0}{1}\otimes \vphi,  f(H(P))\tbinom{0}{1}\otimes\vphi\big\ra 
\end{align*}
for each $\vphi\in \Fock$ with $j\vphi=\vphi$.
\end{Prop}


\begin{thebibliography}{100} 
\bibitem{BFP} V. Bach, J.  Fr\"ohlich, A. Pizzo,  Infrared-finite algorithms in QED: the groundstate of an atom interacting with the quantized radiation field.  Comm. Math. Phys.  {\bf 264},  145--165 (2006)
\bibitem{BFS} V. Bach, J.  Fr\"ohlich, I.  M.  Sigal,   Spectral analysis for systems of atoms and molecules coupled to the quantized radiation field.  Comm. Math. Phys.  {\bf 207},  249--290 (1999)
\bibitem{BCF} V. Bach, T.  Chen, J.  Fr\"ohlich, I. M.  Sigal,  The renormalized electron mass in non-relativistic quantum electrodynamics.  J. Funct. Anal.  
{\bf 243} ,  426--535  (2007) 
\bibitem{Chen}        T.  Chen, Infrared renormalization in non-relativistic QED and scaling criticality
J.   Funct.  Anal.  {\bf 254},   2555-2647 (2008)
  \bibitem{JFroehlich1} J.  Fr\"ohlich,  On the infrared problem in a
  model of scalar electrons and massless, scalar bosons.
  Ann. Inst. H. Poincar\'{e} Sect. A (N.S.) {\bf 19},   1--103 (1973)
  \bibitem{LLG}         M. Griesemer, E. H. Lieb,  M. Loss,  Ground 
        states in non-relativistic quantum electrodynamics, 
        Invent. Math. {\bf 145},   557-595 (2001)   
\bibitem{HH} D. Hasler, I. Herbst,  On the self-adjointness and domain of Pauli-Fierz type Hamiltonians, Rev. Math. Phys. {\bf 20}, 787-800 (2008)
\bibitem{Hiroshima1} F. Hiroshima, Ground states of a model in nonrelativistic quantum electrodynamics. I.  J. Math. Phys.  {\bf 40},  6209--6222 (1999)
\bibitem{Hiroshima2}F.  Hiroshima, Self-adjointness of the Pauli-Fierz Hamiltonian for arbitrary values of coupling constants, Ann. Henri Poincar\'{e} {\bf 3 } 
 171-201 (2002)
\bibitem{Hiroshima3} F. Hiroshima, Fiber Hamiltonians in the non-relativistic quantum electrodynamics, J. Funct. Anal. {\bf 252},  314-355 (2007)
 \bibitem{HS} F. Hiroshima, H. Spohn,  Ground state degeneracy of the Pauli-Fierz Hamiltonian with spin.  
Adv. Theor. Math. Phys.  {\bf 5},   1091--1104 (2001)  
\bibitem{LiebLoss}E. H. Lieb, M. Loss, Existence of atoms and molecules in 
non-relativistic quantum electrodynamics, Adv.  Theor.  
Math.  Phys.   {\bf 7} , 667-710 (2003) 
\bibitem{LMS}         M.  Loss, T. Miyao, H. Spohn,  Lowest energy 
        states in nonrelativistic QED: atoms and ions in motion,  
           J. Funct. Anal. {\bf 243},  353-393 (2007) 
           \bibitem{MS} T. Miyao, H. Spohn, Spectral analysis of the semi-relativistic Pauli-Fierz Hamiltonian,  arXiv:0805.4776 (2008)
                      \bibitem{sasaki}           I. Sasaki,   Ground state of a model in relativistic quantum electrodynamics with a fixed total momentum, arXiv:math-ph/0606029 (2006)
\bibitem{Sigal}  I. M. Sigal,   Ground state and resonances in the standard model of non-relativistic QED,  arXiv:0806.3297 (2008)
\bibitem{Spohn2}  H. Spohn,  Dynamics of Charged Particles and their Radiation Field,
Cambridge University Press 2004
\end{thebibliography}
\end{document}